\documentstyle[prb,aps,multicol]{revtex}
\begin{document}

\title{Theory of Current-Induced Breakdown of the Quantum Hall Effect }
\author{
Kenzo Ishikawa, Nobuki Maeda, Tetsuyuki Ochiai, and Hisao Suzuki}
\address{
Department of Physics, Hokkaido University, Sapporo 060, Japan}
\maketitle
\begin{abstract}
By studying the quantum Hall effect of stationary states with high values 
of injected current using a von Neumann lattice representation, we found 
that broadening of extended state bands due to a Hall electric 
field occurs and causes the breakdown of the quantum Hall effect. 
The Hall conductance agrees with a topological invariant that is 
quantized exactly below a critical field and is not quantized 
above a critical field. 
The critical field is proportional to $B^{3/2}$ and is enhanced 
substantially if the extended states occupy a small fraction of the system. 
\end{abstract}
\draft
\pacs{73.40.Hm, 72.20.My}

The Quantum Hall effect (QHE)\cite{a} is so far the best way of 
measuring a precise value of the fine structure constant experimentally. 
In finite ranges of the electron density and a magnetic field, 
the Hall conductance, $\sigma_{xy}$ stays in a constant value and 
the longitudinal conductance $\sigma_{xx}$ vanishes.\cite{b} 
Width of this quantum Hall regime (QHR), which should be finite 
for the measurement to be possible, is a subject of the present paper. 

In experiments, the injected current is finite. 
The larger current is better for obtaining high precision. 
However it has been found that the width of QHR 
is reduced by increasing injected current and eventually vanishes if the 
current exceeds a critical value. 
The QHE then disappears.\cite{c,d} 
It was suggested by Trugman\cite{da} and Eaves and Sheard\cite{db} 
that electric field causes broadening of Landau levels. 
The Coulomb interaction effect was included by Tsemekhaman et al.\cite{dc} 
In recent experiments, 
Kawaji, Hirakawa, and Nagata\cite{c} found new universal phenomena, 
which can not be understood in previous works, namely the breakdown of the 
QHE occurs if the Hall electric field exceeds the critical value 
which is proportional to $B^{3/2}$ in specially designed butterfly 
shape systems. 

In this paper we develop a theory of the QHE in a finite current system 
and show that mobility edges move and width of 
QHR decreases with Hall electric field. 
If the Hall electric field exceeds the critical value, QHR vanishes. 
The critical field is estimated and is shown to agree with experimental 
results by Kawaji, Hirakawa, and Nagata if the Hall electric 
field for extended states is enhanced by a factor of about twenty. 
The strong localization of electrons due to disorder causes this enhancement. 
We show also that the Hall conductance is quantized exactly and the 
longitudinal conductance vanishes in QHR of systems with small current 
below the critical value.

In a von Neumann lattice representation basis functions 
are localized on the lattice sites. 
Hence this representation is quite useful in studying extended states and 
localized states. We can also obtain exact relations in a strong magnetic 
field with a finite Hall electric field such as the current conservation, 
equal time commutation relations, Ward-Takahashi (WT) identity, 
and a topological formula for the Hall conductance $\sigma_{xy}$. 
A proof of exact quantization of $\sigma_{xy}$ in QHR of the systems with  
disorder and interaction has been given before.\cite{e,f} 
The system with finite injected current is studied in the present paper. 
 
In QHR, one-particle states have an energy gap and stationary current flows 
in a Hall electric field by which Lorentz force is balanced. 
There is no energy dissipation in this system. 
Hence we study a stationary state of a second quantized Hamiltonian that 
has the Hall voltage, $V_H({\bf x})$, and impurity potential 
$V_{\rm imp}({\bf x})$,
\begin{eqnarray}
{\cal H}&=&\int d{\bf x}\Psi^\dagger({\bf x},t)H\Psi({\bf x},t),
\nonumber\\
H&=&{({-i\nabla}+e{\bf A})^2\over 2m}+V_{\rm imp}({\bf x})+
eV_H({\bf x}),\label{eqa}\\ 
&&\partial_x A_y-\partial_y A_x=B.\nonumber
\end{eqnarray}
We use units with $\hbar=c=1$. 
The electron field is expanded by von Neumann coherent states 
$\vert{{\bf R}_{mn}}\rangle$ of guiding center coordinates $(X,Y)$ 
and harmonic oscillator eigenfunctions $\vert f_l\rangle$ of relative 
coordinates $(\xi,\eta)$ with energy eigenvalue $E_l=
\omega_c(l+{1\over2})$, where $\omega_c={eB\over m},\ l=0,1,2,\dots$, 
as,
\begin{eqnarray}
&\Psi({\bf x},t)=\sum a_l({\bf R}_{mn},t)\langle{\bf x}
\vert f_l\otimes{\bf R}_{mn}\rangle,\label{eqb}\\
&(X+iY)\vert{\bf R}_{mn}\rangle=a(m+in)\vert{\bf R}_{mn}\rangle,
\nonumber
\end{eqnarray}
where $m$, $n$ are integers and $a=\sqrt{2\pi/eB}$. 

In Fourier transformed basis of Eq.~(\ref{eqb}), 
which we call an energy basis,\cite{g} 
the free Hamiltonian (not including $V_{\rm imp}$ and $V_H$), 
the current operator, and the commutation relation of the charge density 
are given by
\begin{eqnarray}
&{\cal H}_0=a^2\int{d^2 p\over (2\pi)^2}
\sum b_l^\dagger({\bf p},t)E_l b_l({\bf p},t),\nonumber\\
&j_\mu({\bf k},t)=a^2\int{d^2 p\over (2\pi)^2}\sum b_l^\dagger({\bf p},t)
(\Gamma^{(0)}_\mu({\bf p},{\bf k}))_{ll'} b_{l'}({\bf p}-{\bf k},t),
\label{eqc}\\
&(\Gamma^{(0)}_\mu({\bf p},{\bf k}))_{ll'}=
\langle f_l\vert{1\over2}\{v^\mu,e^{i{\bf k}\cdot{\bf\xi}}\}
\vert f_{l'}\rangle e^{-i{a^2k_x\over4\pi}(2p-k)_y},\nonumber\\
&[j_0({\bf k},t),b_l({\bf p},t)]=-\sum_{l'}b_{l'}
({\bf p}-{\bf k},t)
(\Gamma^{(0)}_0({\bf p},{\bf k}))_{ll'},\nonumber
\end{eqnarray}
where $v^\mu=(1,{\bf v})$, ${\bf v}=\omega_c(-\eta,\xi)$, $\bf k$ is 
the momentum of the continuum two-dimensional space, $\bf p$ is the 
lattice momentum of the von Neumann lattice, and $\vert p_x\vert$, 
$\vert p_y\vert\leq\pi/a$.  
In another basis, which we call a current basis, 
the same quantities are given by 
\begin{eqnarray}
&{\cal H}_0=a^2\int{d^2 p\over (2\pi)^2}
\sum{\tilde b}_{l_1}^\dagger({\bf p},t)(U_{l_1l}({\bf p})
E_l U^\dagger_{ll_2}({\bf p})){\tilde b}_{l_2}({\bf p},t),\nonumber\\
&j_\mu({\bf k},t)=a^2\int{d^2 p\over (2\pi)^2}
\sum{\tilde b}_l^\dagger({\bf p},t)(\tilde\Gamma^{(0)}_\mu({\bf p},
{\bf k}))_{ll'} 
{\tilde b}_l({\bf p}-{\bf k},t),\label{eqd}\\
&{\tilde\Gamma}^{(0)}_\mu({\bf p},{\bf k})=
(1,{\bf v}+{a^2\omega_c\over4\pi}(2{\bf p}-{\bf k})),\nonumber\\
&[j_0({\bf k},t),{\tilde b}_l({\bf p},t)]=
-{\tilde b}_l({\bf p}-{\bf k},t).\nonumber
\end{eqnarray}
Two sets of operators are connected by a unitary transformation,
\begin{equation}
{\tilde b}_l({\bf p},t)=\sum_{l'} U_{ll'}({\bf p}){b}_{l'}({\bf p},t),\ 
U_{ll'}({\bf p})=\langle f_l\vert e^{ip_x\eta}e^{ip_y\xi}\vert f_{l'}\rangle. 
\label{eqe}
\end{equation}
In the energy basis, the free Hamiltonian is diagonal but the current 
operator is non-diagonal in the Landau level index. 
Namely the commutation relation between charge density and 
field operators have level dependent 
higher moment terms, and WT identity becomes complicated. 
In the current basis, on the other hand, the free Hamiltonian is 
non-diagonal but WT identity becomes simple. 
It is convenient to use current basis when we use WT 
identity and related exact relations and to use the energy basis 
when we find energy eigenvalues. 
It is important to notice that there is no representation that 
makes energy and current diagonal at the same time due to the magnetic field. 
This solves a controversy raised by Thouless in Ref.\cite{ga}. 
These properties have not been included in the previous works of 
finite current systems,\cite{da,dc} but play important roles 
in our work. 

Extended states are represented by momentum eigenstates approximately, 
but the localized states are completely different from momentum eigenstates. 
Hence in the system without Hall electric field, the propagator has a momentum 
conserving term due to extended states and a non-conserving term due to 
localized states. 
We can split the propagator in the current basis into a singular part 
proportional to the delta function and regular part as\cite{e}
\begin{equation}
{\tilde S}(p,p')={\tilde S}_c(p)\delta^3(p-p')+
{\tilde S}_n(p,p'),\label{eqf}
\end{equation}
where $p=(p_0,{\bf p})$ and $p_0$ is the angular frequency. 
The first term corresponds to the extended states, and the second term 
corresponds to others. 
The first one contributes to the Hall conductance in 
infinitesimally small current systems. From 
Eq.~(\ref{eqd}), and the current conservation, the momentum 
conserving part of vertex part in the current basis is connected 
with that of the propagator by WT identity,\cite{h}
\begin{equation}
{\tilde\Gamma}_\mu(p,p)={\partial\over\partial p_\mu}{\tilde S}_c
^{-1}(p).\label{eqg}
\end{equation}From Eq.~(\ref{eqg}) the Hall conductance in QHR of small 
current systems becomes a topological invariant of the propagator 
and agrees with an exact integer multiple of $e^2/2\pi$,\cite{e}
\begin{equation}
\sigma_{xy}={e^2\over 2\pi}{1\over24\pi^2}\int{\rm Tr}\{\tilde S_c(p)d
\tilde S^{-1}_c(p)\}^3
\label{eqh}
\end{equation}
Only the momentum conserving term contributes to $\sigma_{xy}$. 
We will see that Eq.~(\ref{eqh}) is valid in a finite current system 
below. 

To study the momentum conserving terms with finite current and a finite Hall 
electric field, we write the Hamiltonian in the energy basis as
\begin{eqnarray}
H&=&H_0+H_1,\nonumber\\
H_0&=&{m\omega_c^2\over2}(\xi^2+\eta^2)+
{eE_H}({a^2p_y\over2\pi}-\xi),\label{eqi}\\
H_1&=&ieE_H {\partial\over\partial p_x}+
\int d^2k V_{\rm imp}({\bf k})\Gamma^{(0)}_0({\bf p},{\bf k}),\nonumber
\end{eqnarray}
where $H_0$ conserves momenta and $H_1$ does not conserve momenta 
and we treat $H_1$ perturbatively.

In the first step, we study the Hall conductance in QHR in finite current 
systems. 
Since the momentum derivative term in $H_1$ generates the derivatives of 
$\delta({\bf p}-{\bf p}')$, the splitting in Eq.~(\ref{eqf}) seems to 
become ambiguous. 
To resolve this difficulty we apply a perturbative expansion to 
the current correlation function 
$\pi_{\mu\nu}(p,p')$ with respect to the momentum derivative 
term and impurity potentials. 
Since the ground state has an energy gap in the lowest order, the 
longitudinal conductance $\sigma_{xx}$ vanishes and the ground state 
is stable. 
The perturbative expansion could be applied as far as the 
Fermi energy is located in QHR. 
$\pi_{\mu\nu}$ includes derivative terms as $\pi^{(d)}_{\mu\nu}
(p,p'){\partial\over\partial p_x}\delta^3(p-p')$, which contributes to 
the total current as ${\partial\over\partial p'_x}\pi^{(d)}_{\mu\nu}(p,p')
\delta^3(p-p')$ by integration by parts. 
Therefore the momentum conserving terms modified by the derivative terms 
contribute to the conductance. 
The Hall conductance is a slope of this modified current correlation 
function at the origin and is written finally as a topological invariant, 
Eq.~(\ref{eqh}), by using the WT identity for the full propagator and 
the full vertex part. As a result ${\tilde S}_c$ has a dependence 
on the Hall electric field $E_H$ and the Hall conductance is quantized 
exactly in QHR of finite current systems. 
Here QHR is defined by a self-energy of the ${\tilde S}_c$ which includes 
Hall electric field.

In the second step we study impurity corrections of the extended state 
energies perturbatively. 
We assume that there are only extended states for the moment. 
Under an assumption for random impurity potentials, 
$
\langle V_{\rm imp}({\bf k}_1)V_{\rm imp}({\bf k}_2)\rangle=
\vert V({\bf k}_1)\vert^2\delta({\bf k}_1+{\bf k}_2)
$, 
the lowest order self-energy correction of the electron within one Landau 
level space is given by
\begin{equation}
\Sigma_{ll}(p)=\sum_{\bf n}\int^{\pi/a}_{-\pi/a}
{d^2k\over(2\pi)^2}{\vert V({\bf p}-{\bf k}-
2\pi{\bf n}/a)\langle f_l\vert e^{i({\bf p}-{\bf k}-2\pi
{\bf n}/a)\cdot{\bf\xi}}\vert f_l\rangle\vert^2\over p_0-E_l-
{eE_H a^2\over2\pi}k_y}-{eE_H a^2\over2\pi}p_y,
\label{eqj}
\end{equation}
where the second term in the right hand side comes from the derivative 
term of $H_1$. 
Note that the Coulomb interaction effect can be expressed in the 
same manner. 
This self-energy has absorptive part in finite energy regions of the 
$l$-independent width,
\begin{equation}
eE_H a.\label{eqk}
\end{equation}
We have examined higher order diagrams which include the interaction 
effect using the self-consistent Born approximation. 
We found that the correction to the $E_H$ dependence of the energy width 
is small and becomes zero in the large $E_H$ limit. 
If the Fermi energy is in these energy regions, Hall conductance is not 
quantized. 
QHR corresponds to the outside of these energy regions.  
                             
We are able to extend the above arguments to systems of finite 
spatial widths. 
The momentum representation is valid also in these systems. 
Momentum in one direction becomes discrete and integration in 
continuum momentum is replaced with summation in discrete momentum. 
Consequently the energy width of extended states in this situation is 
slightly modified but is practically the same as that of infinite system.

So far the uniform Hall electric field is assumed and localized states 
have not been taken into account. 
In real experiments, total current is given and neither current density 
nor Hall electric field are uniform. 
In the last step we estimate the effective Hall field 
$E_H^{\rm eff}$ that the extended states feel in systems with both localized 
states and extended states. 
Localized states, in fact, do not contribute to total current and are 
independent of an injected current. 
Only extended states couple with injected current. 
Conducting charge carrier accumulate toward an edge and cause Hall 
electric field. 
Owing to induced Hall electric field, the injected current flows as a 
stationary current. 
Charges in localized states are concentrated in finite regions and 
do not move long distance. 
Hall electric field is hence connected with only extended states but 
not connected with localized states.
To see this more explicitly, let us study expectation value of the 
time derivative of guiding center coordinates in y-direction,  
\begin{equation}
-i\langle\alpha\vert[Y,H]\vert\alpha\rangle=-
{1\over eB}\langle\alpha\vert{\partial V_H\over\partial X}
\vert\alpha\rangle-{1\over eB}
\langle\alpha\vert{\partial V_{\rm imp}\over\partial X}
\vert\alpha\rangle,\label{eql}
\end{equation}
where $\vert\alpha\rangle$ is an energy eigenstate. 
The left hand side agrees with the velocity and gives electric current of 
the system. 
For the localized states $\vert\alpha,{\rm loc}\rangle$, the left hand 
side vanishes since localized states have normalizable wave functions. 
The localized states carry no current. 
For the extended states $\vert\alpha,{\rm ext}\rangle$, wave functions are 
extended from one edge to another edge and are not normalizable in the 
infinite systems.
Therefore the left hand side can take a non-zero value. 
We define the effective Hall electric field by the right hand side of 
Eq.~(\ref{eql}) and we have
\begin{eqnarray}
&\sum\limits_{\alpha,\rm ext}\langle\alpha,{\rm ext}
\vert v_y\vert\alpha,{\rm ext}\rangle
=\sum\limits_{\alpha,{\rm ext}} {E^{\rm eff}_H\over B} 
\langle\alpha,{\rm ext}\vert\alpha,{\rm ext}\rangle, \nonumber\\
&L_yI_y=e{E^{\rm eff}_H\over B} N_{\rm ext},
\label{eqm}\\
&E^{\rm eff}_H={N_{\rm tot}\over N_{\rm ext}}E^{(0)}_H={1\over\gamma}
E^{(0)}_H,\nonumber
\end{eqnarray}
where $L_x$ and $L_y$ are lengths in x-direction and y-direction, 
$N_{\rm ext}$ is a number of extended states and 
$N_{\rm tot}$ is a number of total states, and 
$E_H^{(0)}$ is an electric field in the absence of localized states, 
$i.e.$, $V_H/L_x$. 
In the above equations, $\gamma$ is a fraction of the extended states, 
and gives an enhancement to the effective Hall electric field. 
If most area of space is occupied by localized states, 
the effective Hall field is enhanced substantially. 
The effects of localized states are included in our theory as an 
enhancement factor $\gamma$ effectively. 

The extended state wave functions are close to free waves and are
modified by impurities. 
Their effects could be included in  perturbative 
treatment  if the enhanced  Hall field is used. We have, finally, 
the following combined width of extended state energies,
\begin{equation}
{1\over\gamma}eE^{(0)}_H a.\label{eqn}
\end{equation}
The energy width of Eq.~(\ref{eqn}) is proportional to $E^{(0)}_H$. 
Outside the energy bands of extended states, there are only 
localized states and edge states. 
Thus QHR is realized in this energy regions. 
The QHR becomes narrower as $E_H$ increases and vanishes when the band 
width equals $\omega_c$ at the critical field $E_c$. 
The extended states cover whole energy regions and the Hall conductance 
is not quantized above $E_c$. 
The critical Hall field is given by,
\begin{equation}
E_c=\gamma{B\over ma}.\label{eqo}
\end{equation}
Thus the critical Hall electric field is proportional to $B^{3/2}$ and 
is enhanced by $\gamma$ if the localized states occupy a 
main part of the total system. 
Actually it has been suggested by Kawaji et al.\cite{i} from their 
experiments of SIMOSFET that the enhancement factor could be order 
twenty. 
If this is of universal value and is valid also in Ga systems,\cite{l} 
our result is consistent with recent experiments.\cite{c} 
More experimental study will be needed to confirm this. 

Estimation of the fraction, $\gamma$, has been made in Ref.\cite{j}.  
These theoretical investigations have been done in one Landau level space. 
They show that the extended states are only in the center of Landau levels 
with vanishing widths. 
This implies that the fraction vanishes and the 
mobility edges do not exist in the thermodynamic limit. 
Our results, however, show that the extended state bands have 
finite energy width and that the 
finite mobility edges exist in realistic systems. 
Inelastic scattering, 
Landau level mixing and other effects which have not been taken into 
account might give this behavior. 
The present work suggested that 
the enhancement factor is universal and of order twenty 
in realistic systems of butterfly shape. 

Our calculations show that there is a clear mobility edge in finite 
current systems. 
In one side of the mobility edge there are two-dimensionally extended 
states and in the other side there are no two-dimensional states. 
It is natural to assume that there are one-dimensional critical 
states at the boundary region of the energy. 
We study an implication of such one-dimensional critical states on 
the phase transition between the Hall liquid and insulator. 
The momentum of the critical states along one direction is 
approximately conserved. 
They do not contribute to Hall conductance but contribute to 
longitudinal conductance. 
Hence the Hall conductance vanishes in this region but 
longitudinal conductance does not vanish. 
Instead, it takes a characteristic value of one dimensional channel. 
The current of one-dimensional channel\cite{ja} is given by,
\begin{equation}
j_y=ev_y n\label{eqp}
\end{equation}
with a velocity $v_y$ and a carrier density near the Fermi surface $n$. 
They are connected with a one-particle energy $E(p)$ through,
\begin{equation}
v_y={\partial E(p)\over\partial p_y},\  
n={1\over2\pi}\Delta p_y,\label{eqq}
\end{equation}
and the chemical potential difference $\Delta\mu$, which is connected 
with the voltage drop $V_y$, 
\begin{equation}
{\partial E(p)\over\partial p_y}\Delta p_y=\Delta\mu=eV_y.
\label{eqr}
\end{equation}
Thus the longitudinal conductance at the phase boundary is given by the 
universal value,
\begin{equation}
J_y={e^2\over2\pi}V_y,\ 
\sigma_{yy}={e^2\over2\pi}.\label{eqs}
\end{equation}
This, in fact, is consistent with the experiments.\cite{k} 

In summary, we have shown that QHR exists in finite current systems and 
the Hall conductance is quantized exactly as integer multiple of $e^2/2\pi$ 
in QHR if the Hall electric field is less than the critical value. 
The extended state bands have finite widths and become wider as the 
Hall electric field increases. 
It exceeds the Landau level spacing at the critical value. 
The critical Hall electric field is proportional to $B^{3/2}$ and 
the proportional constant is enhanced by an amount which depends on the 
fraction of the extended states. 
For the agreement to be obtained compared with the recent 
experiments, about one twentieth of total electrons should be extended, 
which implies that the mobility edges exist at finite energies. 
Critical one-dimensional extended states at the phase boundary between 
localized states regions and extended states regions play important 
roles in the universal conductance at the liquid-insulator transition of 
the quantum Hall systems. 

\acknowledgements

This work was partially supported by the special Grant-in-Aid for 
Promotion of Education and Science in Hokkaido University provided by 
the Ministry of Education, Science, Sports, and Culture, the Grant-in-Aid 
for Scientific Research(07640522), the Grant-in-Aid for Scientific 
Research on Priority area (Physics of CP violation), 
and the Grant-in-aid for International 
Science Research (Joint Research 07044048) from the Ministry of Education, 
Science, Sports and Culture, Japan.


\begin{references}
\bibitem{a}
K. von Klitzing, G. Dorda, and M. Pepper, 
Phys. Rev. Lett. {\bf45}, 494 (1980); 
S. Kawaji and J. Wakabayashi, in {\it Physics in High Magnetic Fields}, 
edited by S. Chikazumi and N. Miura (Springer-Verlag, Berlin, 1981).
\bibitem{b}
H. Aoki and T. Ando, Solid State Commun. {\bf38}, 1079 (1981).
\bibitem{c}
S. Kawaji, K. Hirakawa, and M. Nagata, Physica B {\bf 184}, 17 (1993); 
T. Okuno et al., J. Phys. Soc. Jpn. {\bf 64}, 1881 (1995). 
\bibitem{d}
G. Ebert et al., J. Phys. C16, 5441 (1983); M. E. Cage et al., 
Phys. Rev. Lett. {\bf51}, 1374 (1983); 
see also, N. Balaban, U. Meirav, and H. Shfrikman, 
Phys. Rev. {\bf B52}, R5503 (1995).
\bibitem{da}
S. Trugman, Phys. Rev. {\bf B27}, 7539 (1983). 
\bibitem{db}
L. Eaves and F. W. Sheard, Semicon. Sci. Tech. {\bf1}, 346 (1986). 
\bibitem{dc}
V. Tsemekhaman et al., Phys. Rev. {\bf B55}, R10201 (1997).
\bibitem{e}
N. Imai, K. Ishikawa, T. Matsuyama, and I. Tanaka, Phys. Rev. 
{\bf B42}, 10610 (1990); K. Ishikawa, Prog. Theor. Phys. Suppl. 
{\bf 107}, 167 (1992).
\bibitem{f}
K. Ishikawa, N. Maeda, and K. Tadaki, Phys. Rev. {\bf B54}, 17819 (1996).
\bibitem{g}
K. Ishikawa, N. Maeda, T. Ochiai, and H. Suzuki, Phys. Rev. {\bf B58}, 
1088 (1988); cond-mat/9806185. 
\bibitem{ga}
D. J. Thouless, J. Phys. {\bf C17}, L325 (1984); I. Dana and J. Zak, 
Phys. Rev. {\bf B28}, 811 (1983); E. Brown, Phys. Rev. {\bf133}, 
1038 (1964).
\bibitem{h}
J. C. Ward, Phys. Rev. {\bf 78}, 1824 (1950); Y. Takahashi, Nuovo Cimento 
{\bf6}, 370 (1957).
\bibitem{i}
S. Kawaji, J. Wakabayashi, and J. Moriyama, J. Phys. Soc. Jpn. {\bf50}, 
3839 (1981).
\bibitem{l}
M. Furlan, Phys. Rev. B{\bf57}, 14818 (1998).
\bibitem{j}
Y. Huo and R. N. Bhatt, Phys. Rev. Lett. {\bf68}, 1375 (1992); 
Y. Ono, J. Phys. Soc. Jpn, {\bf51}, 2055 (1982); 
T. Ando and H. Aoki, J. Phys. Soc. Jpn. {\bf54}, 2238 (1985); 
B. Huckestein and B. Krammer, Phys. Rev. Lett. {\bf64}, 1437 (1990). 
\bibitem{ja}
M. B\"uttiker, Phys. Rev. {\bf B38}, 9375 (1988); 
R. Landauer, IBM J. Res. Dev. {\bf 1}, 223 (1957).
\bibitem{k}
D. Shahar et al., Phys. Rev. Lett. {\bf74}, 4511 (1995).
\end{references}
\end{document}